\documentclass[fleqn,usenatbib,useAMS,letters]{mnras}

\usepackage{graphicx}	
\usepackage{amsmath}	
\usepackage{amssymb}	
\usepackage{multicol}        
\usepackage{bm}		
\usepackage{pdflscape}	
\usepackage{comment}	

\usepackage[utf8]{inputenc}

\newcommand{\kms}{\,km\,s$^{-1}$} 
\newcommand{\ergs}{\,ergs\,s$^{-1}$} 

\usepackage[T1]{fontenc}
\usepackage{ae,aecompl}

\usepackage{newtxtext,newtxmath}

\title[A 2 day orbital period for a redback MSP Candidate in the Globular Cluster NGC~6397]{A 2 day orbital period for a redback millisecond pulsar candidate in the globular cluster NGC~6397}

\author[Manuel Pichardo Marcano, other]{Manuel Pichardo Marcano$^{1}$, \thanks{Contact e-mail: \href{manuel.pichardo-marcano@ttu.edu}{manuel.pichardo-marcano@ttu.edu}}{L.E. Rivera Sandoval$^{1,2}$, Thomas J.  Maccarone $^{1}$, Yue Zhao$^{2}$, }
\newauthor{Craig O. Heinke$^{2}$}
\\
$^{1}$Department of Physics and Astronomy, Texas Tech University, Lubbock, TX 79409\\
$^{2}$Department of Physics, University of Alberta, CCIS 4-183, Edmonton, AB, T6G 2E1, Canada}

\date{-}

\pubyear{2019}
\hypersetup{draft}
\begin{document}
\label{firstpage}
\pagerange{\pageref{firstpage}--\pageref{lastpage}}
\maketitle

\begin{abstract}
We report optical modulation of the companion to the 
X-ray source 
U18 in the globular cluster NGC 6397. U18, with combined evidence from radio and X-ray measurements, is a strong candidate as the second redback in this cluster, initially missed in pulsar searches. This object is a bright variable star with an anomalous red color and optical variability ($\sim0.2$ mag in amplitude) with a periodicity $\sim 1.96$ days that can be interpreted as the orbital period. This value corresponds to the longest orbital period for known redback candidates and confirmed systems in Galactic globular clusters and one of the few with a period longer than 1 day. 
\end{abstract}





\section{Introduction}

Redback pulsars are a sub-type of binary millisecond pulsars (MSPs) with main sequence, or slightly evolved, low mass companions ($M_{c} \simeq 0.2 - 0.9 M_\odot$) \citep{MalloryRedback2013}. They differ from the closely related black widows by having companions which are non-degenerate, and a bit more massive, than the $\lesssim0.1 M_\odot$  degenerate companions in the black widow systems. Both classes are characterized by radio eclipses due to excess ionized gas within the system, which suggests that the companion is being ablated by irradiation from the pulsar wind \citep{FruchterEclipse1988,Fruchter88b}. These systems are important for understanding the evolution of MSPs as their formation mechanisms and eventual fates are open questions.

\cite{ChenFormationRedback2013} and \cite{Benvenuto2014} propose contrasting views about the evolutionary tracks leading to the formation of a redback system, and whether or not these end as black widows. 
\cite{ChenFormationRedback2013} concluded that redbacks and  black  widow  pulsars  follow  different  evolutionary  paths depending  on  the  strength  of  evaporation,  finding  that  black widows  do  not  descend  from  redback  pulsars, while \cite{Benvenuto2014} concluded that black widows can descend from redbacks, depending on the evaporation  and  irradiation  feedback on the system. 

Other open questions remain regarding redbacks and their relation with transitional MSPs (tMSPs).\footnote{tMSPs are accreting millisecond pulsars that switch between accreting X-ray binary and millisecond radio pulsar states.} A handful of 
redbacks are 
 confirmed tMSPs (PSR J1023+0038, PSR J1824-2452I, and PSR J1227-4853;  \citealt{Archibal09,Papitto13,bassa2014change}).  It is not clear if all redbacks will eventually show transitional behaviour if observed over a long period of time, but long-term follow up of these systems is vital and can help guide the needed theoretical work on this topic
.

Optical and X-ray observations can provide relevant information regarding the redback systems and help guide future radio searches. The X-ray spectra of redbacks consist of a dominant non-thermal component and at least one thermal component, likely originating from the heated pulsar polar caps \citep{Zavlin96,Zavlin02,Bogdanov05,Bogdanov2011Pulsar}. The non-thermal emission from MSPs can be produced by two different mechanisms: coherent emission from the magnetosphere or non-thermal emission due to an intrabinary shock between the pulsar wind and the companion's material \citep{2003SciStapper}. The latter has been shown to be important for the confirmed redback MSPs, as the non-thermal emission is orbitally modulated \citep{Bogdanov05,bogdanov_chandra_2010,Bogdanov2011Pulsar}. In the optical, redbacks show orbital modulation due to effects such as the ellipsoidal modulation of the near-Roche filling companion, and day vs night side temperature differentials due to heating of the secondary from the pulsar or intrabinary shocks \cite[e.g.][]{Callanan95,Romani16,Cho2018}. Modeling of their optical lightcurves has been done for a handful of these systems, to obtain constraints on the component masses and  inclinations \citep{Thorstensen2005,Cho2018,Strader2019Redbacks}.


The known redbacks are also of special interest as 
they 
have higher average neutron star masses than the canonical mass of $\sim 1.4M_\odot$, with a median mass of $1.78 \pm 0.09 M_{\odot}$ \citep{Strader2019Redbacks}, with some having evidence of hosting neutron stars with masses larger than $2 M_\odot$ \citep{Linares2018,Strader2019Redbacks,Kandel20}. Mapping out their populations and evolutionary tracks is thus important for understanding binary evolution. Unfortunately, these systems can be particularly hard to find in radio pulsar searches as the pulses from the neutron star can be eclipsed, or scattered out, by the ionized stripped gas from the hydrogen rich companion star \cite[e.g.][]{FruchterEclipse1988}. Currently there are 25 redbacks or redback candidates in the Galactic field \citep{Strader2019Redbacks,2020SwihartRedback}, and 
13 confirmed redbacks in 8 globular clusters. These pulsar searches can be computationally demanding, especially for an unknown orbital period. Furthermore, the variation in the dispersion measure expected in the redback systems as the radio pulse propagates through the companion's wind can add an additional computational parameter 
to the search. 




NGC 6397 is the closest (2.4 kpc) core-collapsed globular cluster,  
  with an average reddening of $E(B - V) = 0.18$ \citep{harris_catalog_1996,mclaughlin_resolved_2005}. Due to its proximity and low reddening, NGC 6397 has been extensively studied at different wavelengths. 
  \cite{cool_discovery_1993} first detected X-ray sources in NGC 6397, with the {\it ROSAT} satellite. 
  Optical and UV observations with the {\it Hubble Space Telescope} (HST) have enabled the identification of several X-ray sources with blue, variable, and/or H$\alpha$ bright stars \citep{cool_discovery_1995,grindlay_spectroscopic_1995,Cool98,taylor_helium_2001,Ferraro01,Shara05}. Identifications were dramatically improved by the  sub-arcsecond positions of the {\it Chandra X-ray Observatory} \citep{grindlay_chandra_2001,bogdanov_chandra_2010}, enabling identification of many more optical/UV counterparts with HST \citep{kaluzny_photometric_2006,cohn_identification_2010,Heinke14,Pallanca2017Halpha,Dieball2017UV}.
NGC~6397 has also been observed with radio telescopes, both imaging observations with the {\it Australia Telescope Compact Array} ({\it ATCA}) \citep{CoryPaper}, and 
radio pulsar searches 
with Parkes \citep{damico2001discovery} that 
discovered PSR J1740-5340  (aka the X-ray source U12, \citealt{grindlay_chandra_2001}) and confirmed its nature as an eclipsing redback system.

Using ground-based data, \cite{kaluzny_photometric_2006} reported a 1.3 day periodicity for the source V31, which they associated with U18. 
\cite{bogdanov_chandra_2010} studied the X-ray properties of U18 and found similarities in its luminosity and spectrum with the confirmed MSP U12, as both sources have a dominant non-thermal component, possibly from  an intrabinary shock.  \cite{bogdanov_chandra_2010} did not find the claimed periodicity of 1.3 days reported by \cite{kaluzny_photometric_2006} in the X-ray data of U18, and dismissed V31 as the possible optical counterpart of U18 based on the distance between them ($~2.1 \sigma$) and the presence of a more likely optical counterpart closer to the X-ray source \citep{grindlay_chandra_2001}. 
 \cite{bogdanov_chandra_2010} suggested U18 as a likely second MSP in the cluster, based on the similarity of its X-ray and optical properties to those of the MSP PSR J1740-5340. 
 This interpretation of U18 as a redback was further strengthened by 
the radio detection of U18 with a steep spectrum, as typical of pulsars 
\citep{CoryPaper}.

U18 has also been  observed twice with the Multi Unit Spectroscopic Explorer  \citep[MUSE][]{MUSECommissioning}. \cite{Husser2016MUSE}  reported two observations 22 h apart, during MUSE commissioning in Summer 2014. In their full-spectrum fitting analysis they found a K subgiant with  $\log g \approx 3.6$ and $T_{\text{eff}} \approx 4200 K$ as the best fit for the optical counterpart of U18, and calculated a radial velocity difference between the two observations of $ \Delta  v_{\text{rad}} =   145.9$ \kms.

Here we 
discover an orbital period of $1.96 \pm 0.06$ days for the strong redback candidate U18.   As U18 is in one of the closest globular clusters, 
it is a perfect candidate for high resolution spectroscopic follow-up to fully characterize its orbit, adding to a small sample of redbacks with well constrained dynamically determined masses.

\section{Data Analysis and Reduction}


We use WFPC2 data from the parallel field of the HST large program 
GO-10424 \citep{2006Sci...313..936R}, which covered the cluster center (the primary field, using the ACS camera, is arcminutes away). The data consist of 126 orbits, and each WFPC2 orbit is divided into three exposures, in the filters F814W, F606W and F336W.
We use the F336W data to avoid crowding in the field, taking advantage both of the somewhat better diffraction limit in the bluer band, and the smaller number of bright stars in the bluer band given the globular cluster's old stellar population. This results in 126 individual exposures (one exposure per orbit) in F336W with exposure times ranging from 500-700 seconds, taken between mid-March and early April 2005 (2005-13-03 to 2005-08-04). The minimum separation between data points being 74 minutes and the maximum of 3.2 days with a total baseline of 26 days.

 \begin{figure}
	\includegraphics[width= 0.9 \columnwidth]{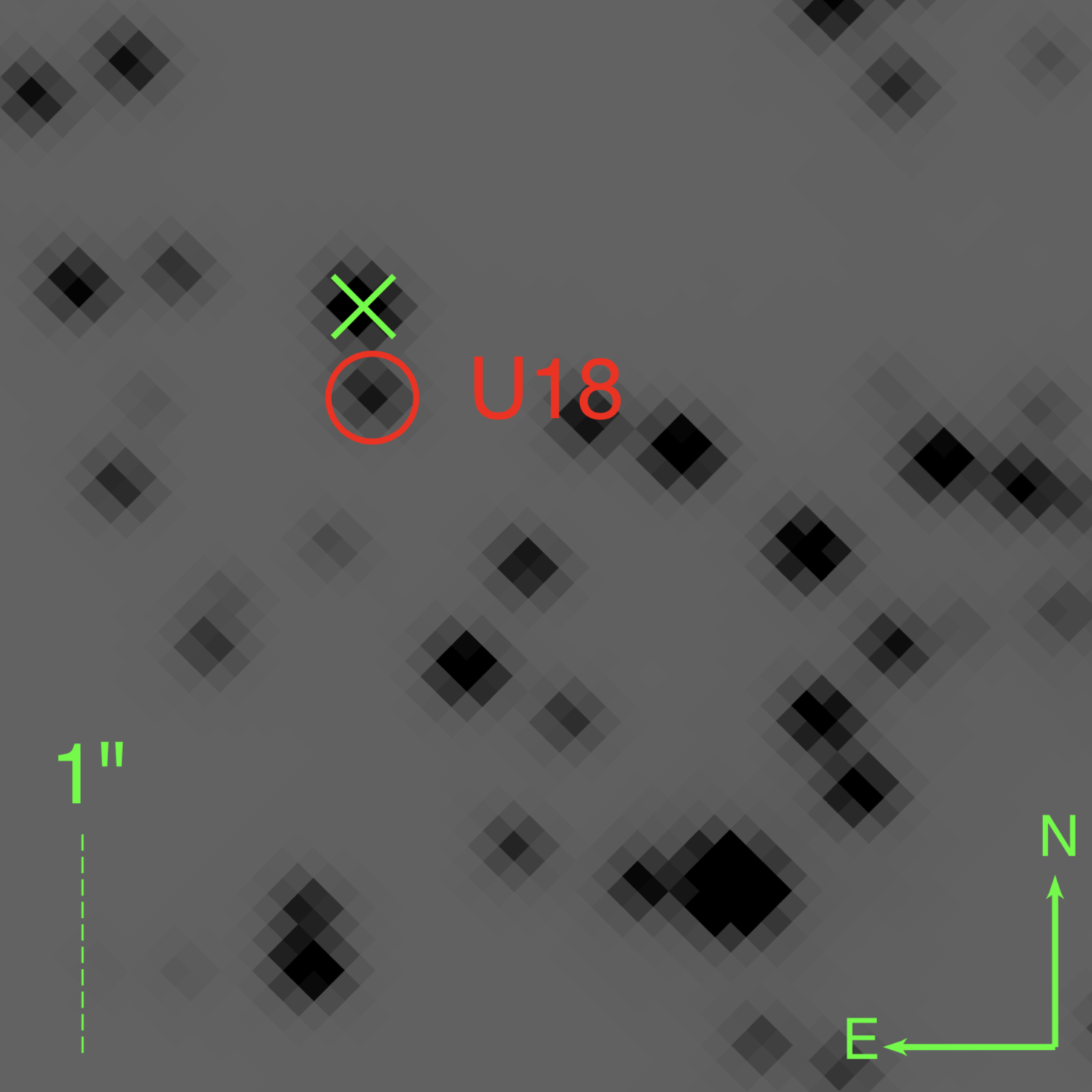}
    \caption{5"$\times$5" WFPC2 image around U18 in F336W. U18 is marked by the 0.2" radius red circle. V31 is the star north of U18 $\sim 0.4"$ away marked with an X. North is up and East is left.}
    \label{fig:u18}
\end{figure}

 
 For the photometry, we use DOLPHOT \citep{2000PASP..112.1383D} including the WFPC2 Module to perform point-spread-function photometry. We supply DOLPHOT the calibrated single-exposure image data (c0m) WFPC2 files and the drizzled .drz image as the reference frame for alignment. The final output from the software lists the position of each star relative to the reference image, together with the measured aperture-corrected magnitudes calibrated to the Vega system for the individual exposures, along with some diagnostic values. We limit the data to measurements containing an error flag of zero meaning that the star was recovered extremely well in the image without contamination due to cosmic rays.

\begin{figure}
	\includegraphics[width=\columnwidth]{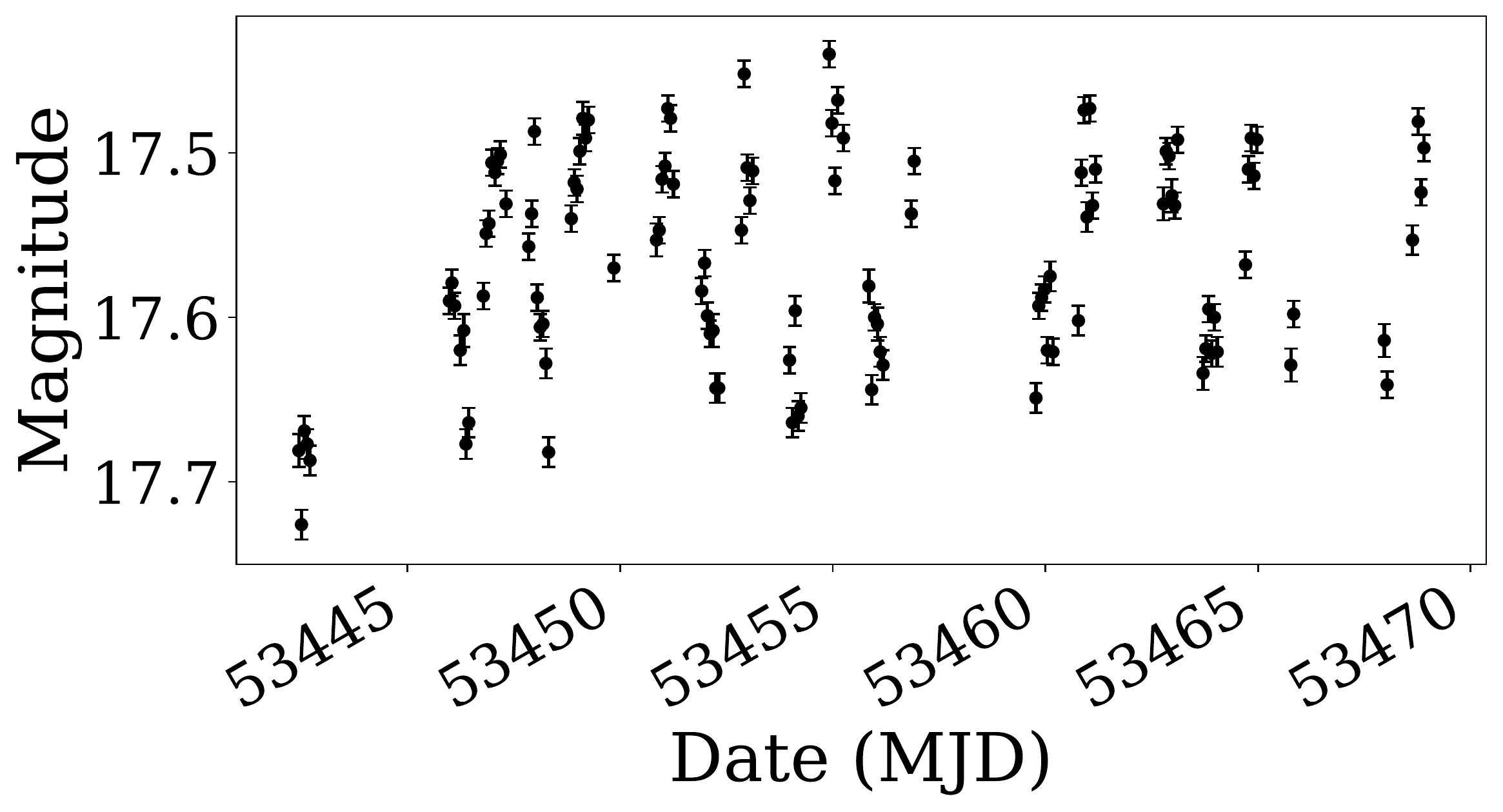}
    \caption{Lightcurve for the redback millisecond pulsar candidate, U18. The magnitudes are from the F336W filter without dereddening and magnitudes uncertainty from the DOLPHOT output catalog.}
    \label{fig:lc}
\end{figure}

\begin{figure*}
	\includegraphics[width=0.6\textwidth]{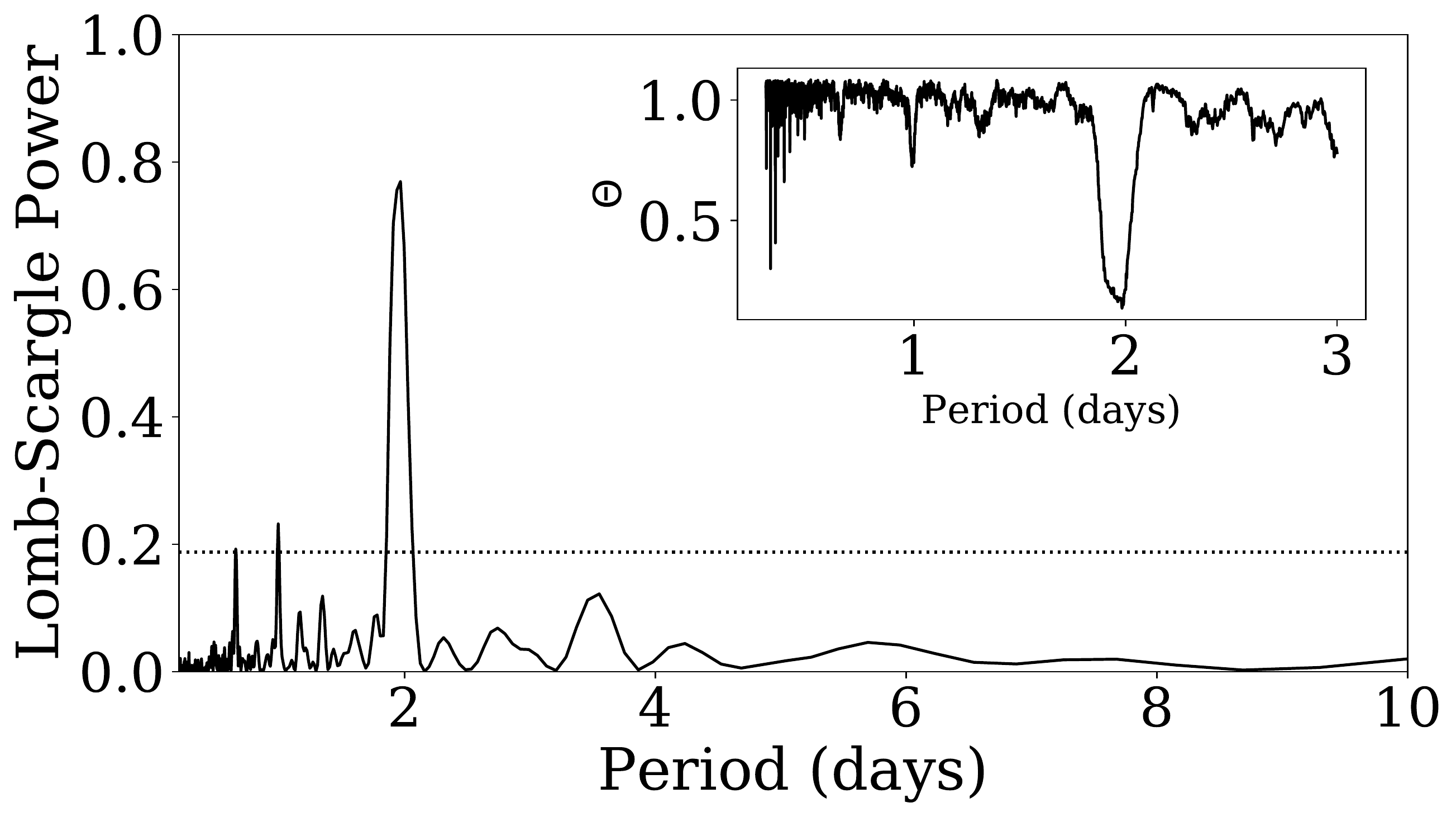}
    \caption{Lomb-Scargle Periodogram for 
    U18, showing a clear isolated peak around 1.96 days. The dotted line shows the periodogram level corresponding to a maximum peak false alarm probability of $1\%$, using the bootstrap method that simulates data at the same observation times to approximate the true distribution of peak maxima for the case with no periodic signal present. The bootstrap method is performed with periods between 0.1 and 10 days, and after normalizing by the residuals of the data around a constant reference model. The inset on the graph shows the $\Theta$ statistics as defined by \protect\cite{PDMStellingwerf78}. The period that minimizes the dispersion of the data at constant phase is $\sim 1.98$ days. }
    \label{fig:periodogram}
\end{figure*}

\begin{figure}
	\includegraphics[width =\columnwidth]{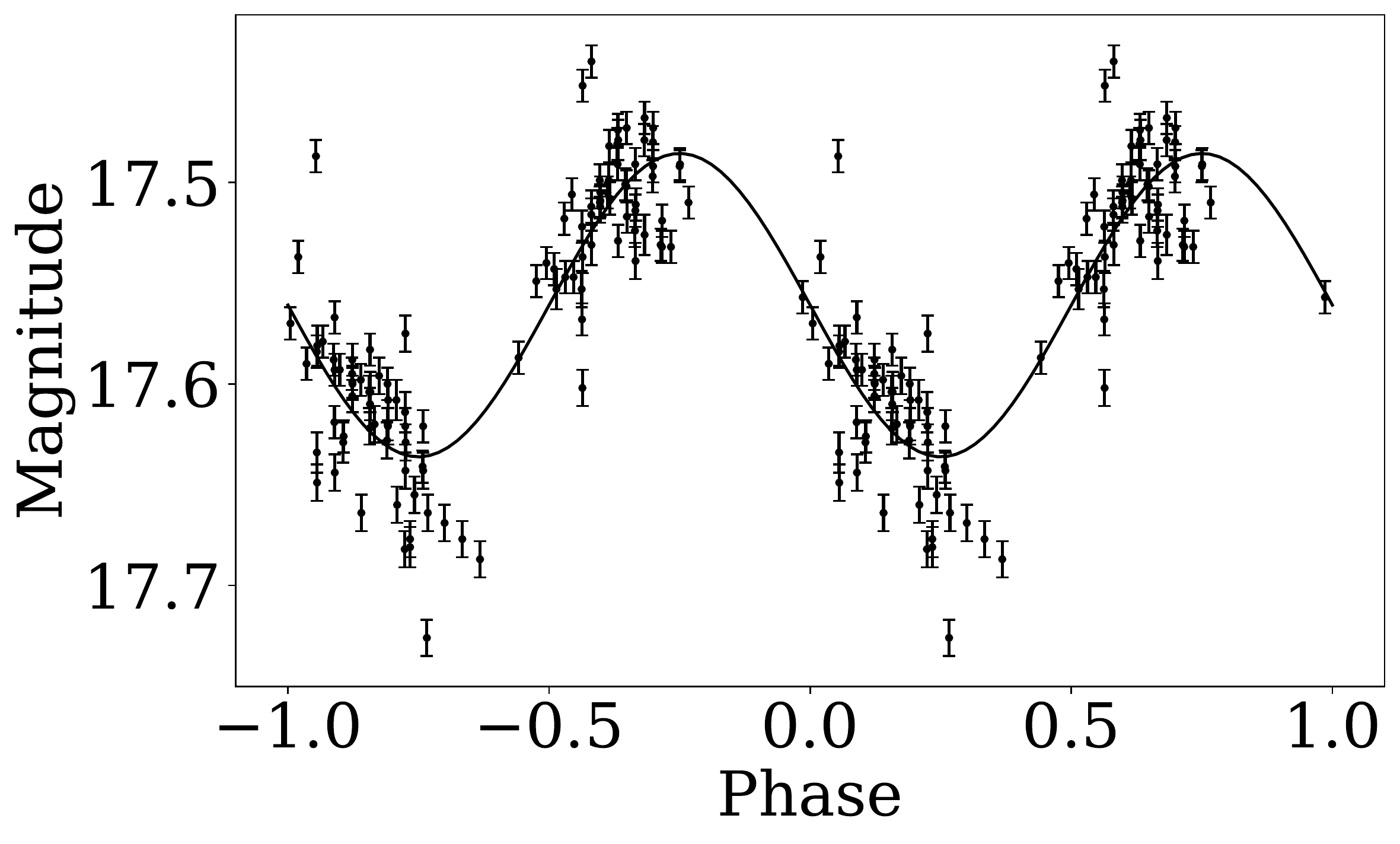}
	\vspace{-0.5 cm}
    \caption{Lightcurve for U18 folded on the 1.96 day period. The magnitudes are from the F336W filter without dereddening. }
    \label{fig:folded}
\end{figure}

 
From the DOLPHOT output, we construct a lightcurve for the optical counterpart of U18 (see Fig.~\ref{fig:u18}). The clean data set with well-measured magnitudes from the software consists of 112 data points (Fig.~\ref{fig:lc}). 
We then perform a period search using the Lomb-Scargle \citep{1976Ap&SS..39..447L,1982ApJ...263..835S} periodogram. The periodogram is then normalized by the residuals of the data around a constant reference model. The periodogram 
(Fig.~\ref{fig:periodogram}) clearly shows an isolated peak at a period of $1.96$ days, the precision with which this peak’s frequency can be identified is directly related to the width of this peak. For the uncertainty in the period we take the $\sigma$ of the Gaussian fit of the peak $\sim 0.06$ day. We also calculate a false alarm probability of $6.6\times 10^{-32}$ using the method described in \cite{Baluev}. The folded lightcurve at that period is shown in Fig.~\ref{fig:folded}.

We also perform a periodicity analysis using the phase dispersion minimization technique  \citep[PDM][]{PDMStellingwerf78}. In this method, the lightcurve is divided into different phase bins and the cost function, $\Theta =s^2/\sigma^2$, is minimized to choose the best period. 
Here $s$ is the phase bin variance and $\sigma$ is the total data variance. We divide the data into 10 equidistant bins to calculate $\Theta$. The plot of the ratio $\Theta$ vs. period (inset in Fig.~\ref{fig:periodogram}) is a good indicator for the best  period. 
The local minimum corresponds to $\Theta = 0.1365$ at a period of 1.98 days, consistent with the results obtained with the Lomb-Scargle periodogram. To calculate the false alarm probability (FAP), an estimate of the significance of the minimum against the hypothesis of random noise, \cite{pdmbeta97} showed that the PDM statistic follows a beta distribution and thus the FAP can be calculated as:

$$\text{FAP} = 1-\left [ 1 - \beta\left ( \frac{N-M}{2},\frac{M-1}{2},\frac{(N-M)\times \Theta}{N-1}\right )\right ]^m$$

\noindent where N is the total number of data points, M is the number of bins, $\Theta$ is the value of the PDM theta statistic, and $m$ is the effective number of independent frequencies. The number of independent frequencies can be estimated in a number of ways \citep[e.g.,][]{Nemec1985,HorneLombFreq1986,Cumming2004}, here we follow the conservative prescription given in \cite{GuidePeriodSearch2003}  and choose $m= \min(N, N_f,\Delta f \Delta T)$ where $N_f$ is the number of frequencies, $\Delta f$ is the frequency range and $\Delta T$ is the time span of observations. This gives a very small FAP of  $5\times 10 ^{-40}$.

Red noise 
can produce spurious detections of periodic behaviour in datasets with small numbers of cycles of the putative period \citep{Press1978RedNoise}.
 To test against this possibility, we simulate random red noise light curves following the algorithm detailed in \citet{Timmer1995Rednoise}. The algorithm randomizes the phases and the amplitudes of an underlying power law spectrum by drawing two sets of normally distributed numbers, and then inverse Fourier transforming them. First, we produce a simulated light curve that exhibits a power law spectrum of the form $(1/f)^2$ evenly sampled at the exposure time of our observations ($\sim 10$ min). Then, we add random Gaussian noise with the same mean and variance as the errors from the observations. The next step is to re-sample the  simulated light curve with noise added and  re-scale it to have the same sampling characteristics, mean and variance as the light curve of the observation. For each generated mock light curve, we compare the maximum power in the Lomb-Scargle periodogram to that of the observed periodicity. We repeat this step $10^6$ times to estimate the probability that the found periodicity is due to underlying red noise. In our $10^6$ trials, only 10 periodograms had a higher power than the one found in the real lightcurve. This corresponds to a probability of $0.001\%$.  In all cases, the spurious periods were 8-10 days, much longer than the real one. Our data spans a bit more than 13 period cycles, and fig.~\ref{fig:periodogram} clearly show only weak variability on longer timescales.  In combination with the other indications that the system should have an orbital period of about 2 days, discussed below, the number of observed cycles and our simulated lightcurves present a strong case against a spurious period due to red noise.


 \section{Results and Discussion}

 
 The strong optical modulations at 1.96$\pm$0.06 days can be explained by the heating of one side of the companion by radiation from the pulsar, as seen in other confirmed redbacks in the field \citep[e.g.,][]{Hui2015PSR2219,salvetti2015J2039}.  This interpretation is further strengthened by the recently reported steep radio spectrum of U18 \citep{CoryPaper}. \citet{CoryPaper} report an spectral flux density at 5.5 GHz of $S_\nu =54.7 \mu\text{Jy}$. A 1.96 day orbit would put U18 as the longest period confirmed redback in a globular cluster, and one of the few redbacks with an orbital period longer than 1 day in the Galactic field. Two other redbacks in the field, 1FGL J1417.7-4407 \citep{Strader2015,Camilo2016,Swihart2018} and 2FGLJ0846.0+2820 \citep{Swihart2017} have longer periods at 5.3 and 8.3 day respectively. 

 The strong radio emission also argues against an active binary scenario. \citet{Guedel1993} found that for coronally active stars, typically $\log(L_R/L_X) \sim -15.5 \text{ Hz}^{-1}$. From this relationship, for U18, assuming a $L_X= 6.7 \times 10 ^{31}$\ergs (0.3-8 keV) \citep{bogdanov_chandra_2010}, $L_R = 2 \times 10^{16}$ \ergs $\text{ Hz}^{-1}$. This is an order of magnitude lower than the reported value for $L_R (5.5 GHz) = 3.8\times 10^{17}$ \ergs $\text{ Hz}^{-1}$ reported by \citet{CoryPaper}.

 The optical to X-ray luminosity argues further against the active binary scenario. \cite{Pallanca2017Halpha} calculate the dereddened  magnitude for U18 as $V_0 = 16.05$, from the HST F555W observations of the cluster. We take a distance modulus $(m-M)_0 = 12.01$ \citep{Gratton2003} and a $T_{eff} = 4200 \, K$ \citep{Husser2016MUSE}. We can use the \cite{Flower1996BC} bolometric correction to find a bolometric luminosity, $L_{opt} = 1.6 \times 10^{34}$ \ergs.  Then, with $L_X= 6.7 \times 10 ^{31}$\ergs \citep[0.3-8 keV;][]{bogdanov_chandra_2010}, we determine that the ratio is $L_X/L_{opt} = 4.2 \times 10^{-3}$, well above the saturation limit of $10^{-3}$ for active binaries \citep{VilhuSaturation2,VilhuSaturationLimit}, arguing strongly against that scenario. We can further exclude the possibility that the X-rays were measured during a flaring event. Large flares, with total energies of order $10^{37}$\ergs  (0.1-2.4 keV) have been detected in active binary systems, like RS CVn \citep[e.g.,][]{Kuerster1996}. U18 has been observed in five different epochs from 2000 to 2007 for a total of 340 ks in the X-ray with Chandra. \citet{bogdanov_chandra_2010} analyzed all 5 observations and reported similar count rates for all of them, making it less likely that U18 was caught during a flaring event. 
Additionally, using the same data sets than \citet{bogdanov_chandra_2010}, we also searched for candidate periods in the archive Chandra X-ray data, but the sampling of the data is not ideal to find a period in the order of days, and we did not find any convincing periodicity in the X-rays (not even folding at the optical period we identified).

 The radio emission also makes the identification as a CV less likely. Furthermore, \cite{cohn_identification_2010} examined the ratio of X-ray to optical flux for many optical counterparts of Chandra X-ray sources. They conclude that U18 has a similar ratio to the confirmed MSP U12 and separated from other CV candidates.Also  \cite{Pallanca2017Halpha} looked at the equivalent width of the H$\alpha$ emission of U18 and measured an equivalent width similar to U12 and smaller than other confirmed CV and CV candidates in the cluster. All this argues against U18 being a CV candidate.
 
 We can use the proposed orbital period to find some constraints on the properties of U18. From the obtained orbital period and the X-ray luminosity, we can place a limit on the spin-down luminosity. We use the relation from \cite{Possenti2002} $$L_x = 10^{-15.3} \times L_{sd}^{1.34} $$ where $L_x$ is the X-ray luminosity from 2-10 KeV and $L_{sd}$ is the spin-down luminosity. A $L_X= 6.7 \times 10 ^{31}$\ergs \citep[0.3-8 keV;][]{bogdanov_chandra_2010} gives a $L_{sd} = 1.47\times 10 ^{35}$ \ergs. Assuming that the system is close to being Roche-lobe filling and using the relationship from \cite{EggletonRoche1983}, for a mass range of $q\sim0.1-1$, the companion would receive a total power output on the order of $\times 10^{33}$ \ergs. This corresponds to $\sim 10 \%$ of the total bolometric luminosity of the star. Relating this power to the $\Delta mag$ observed for this system requires more data in other wavelengths to do an accurate modelling taking into account the albedo and the fraction of light from intrabinary shocks in the system. Furthermore, the light curve is in $F336W$ or near UV where in other systems it has been shown that there is variability due to intrabinary shocks \cite[e.g.][]{Liliana201847Tuc}. In other systems where no ellipsoidal variations are observed the magnitude of the modulations decreases toward longer wavelengths \cite[e.g.][]{Baglio2016}. U18 is near the core of the cluster and only $\sim 0.2"$ from a brighter star. Due to extreme crowding at redder wavelengths and possibly lower amplitude of modulations, we limit the variability study to the F336W data. Future data with adaptive optics or speckle interferometry to study the variability of this source at redder bands would allow to do a detailed modeling of the system. 
 
Other evidence that points to intrabinary shocks as the source of the blue light is the observed colors for U18. The V-I colors of this object, $V-I = 0.93$ \citep{Pallanca2017Halpha}, are significantly bluer than expected for a $T_{eff}=4200$ K star \citep{Mamajek2013} (which is indicated from spectroscopy), indicating that there is an extra component providing blue continuum light. This means that the observed flux is a combination of emission from the companion star and from the intrabinary shock, making the companion to look brighter and ’bluer’ that in reality.

 We also compared the observed limits on the size of the donor star with predictions for nearly Roche-lobe filling companions at different orbital periods in order to test whether the 1.96 day period is the actual orbital period or a harmonic of the orbital period (as might be expected if the modulations are ellipsoidal). In redder bands we expect light to dominates by the companion. For a star of $T_{eff} \approx 4200 \, K$, the theoretical radius is $R_\odot = 0.676 R_\odot$ and the absolute magnitude in the I band  $M_{I} = 6.2$ \citep{Mamajek2013}. Using the extinction from \cite{Richer2008}, this gives a theoretical magnitude in the I band of $M_I = 5.87$. We can compare this to the reported values in the I band for U18. From \cite{Pallanca2017Halpha} the absolute magnitude of U18 is  $M_{I_o} = 3.210$, comparing these two values, we can infer that the companion is 11.5 times brighter than a theoretical star at the same $T_{eff}$, and thus has a radius $R \approx 2.3 R_\odot$. Assuming typical values for the mass ration $q=0.16$ and the mass of the neutron star, $M_{NS}=1.76$ \citep{Strader2019Redbacks}. For the proposed orbital period of 1.96 days, assuming a Roche-lobe filling companion, as is the case in many redbacks, and using the \cite{Paczy1971} relations for the radius of a Roche-lobe star and Kepler's second law we get an estimated radius of the companion of $R_{companion} = 2 R_\odot$. This is close to the estimated value of the radius from the measured I band flux of U18. A longer orbital period of $P_{orb}=2\times 1.96$ days gives a $R_{companion} = 3.2 R_{\odot}$. To get a $R_{companion} = 2 R_\odot$ with twice the proposed orbital period, the mass ratio would need to be low compare to other known Redbacks (\cite{Strader2019Redbacks} reports a minimum $q= 0.07$) and would make the companion a star with $M =0.07 M_\odot$ and $R= 2 R_\odot$, making it a system with an unusually low mass ratio. All this argues for the period to be 1.96 days and argues against a longer orbital period.

\section{Acknowledgement}

This research made use of Astropy \citep{astropy:2013, astropy:2018} and Matplotlib \citep{matplotlib}.

\section{Data Availability}

This work used public data available in the Mikulski Archive for Space Telescopes under the programme GO-10424 (P.I.H.Richer).


\bibliographystyle{mnras}
\bibliography{redback} 

\bsp	
\label{lastpage}
\end{document}